\documentclass[11pt]{article}

\usepackage[margin=1in]{geometry}
\usepackage{amsmath,amssymb,amsthm,mathtools}
\usepackage[T1]{fontenc}
\usepackage{lmodern}
\usepackage{microtype}
\usepackage{enumitem}
\usepackage{booktabs}
\usepackage{array}
\usepackage{algorithm}
\usepackage[noend]{algpseudocode}
\usepackage{xcolor}
\newif\ifshowauditchanges
\showauditchangesfalse
\newcommand{\auditrevision}[1]{%
  \ifshowauditchanges{\color{blue}#1}\else#1\fi}

\newif\ifshowchanges
\showchangesfalse
\ifshowchanges\colorlet{changes}{blue}\else\colorlet{changes}{black}\fi
\usepackage{tikz}
\usetikzlibrary{arrows.meta,decorations.markings,calc}
\usepackage[colorlinks=true,linkcolor=blue!50!black,citecolor=blue!50!black,urlcolor=blue!50!black]{hyperref}

\newtheorem{theorem}{Theorem}[section]
\newtheorem{lemma}[theorem]{Lemma}
\newtheorem{corollary}[theorem]{Corollary}

\theoremstyle{definition}

\theoremstyle{remark}

\newcommand{\Z}{\mathbb{Z}}
\newcommand{\cC}{\mathcal{C}}
\newcommand{\cS}{\mathcal{S}}
\newcommand{\Med}{\operatorname{Med}}
\newcommand{\finf}{f_\infty}
\newcommand{\eps}{\epsilon}
\newcommand{\TV}{\mathrm{TV}}
\newcommand{\E}{\mathbb{E}}
\newcommand{\Ham}{\mathrm{Ham}}
\newcommand{\one}{\mathbf{1}}
\newcommand{\cE}{\mathcal{E}}
\newcommand{\tmix}{t_{\mathrm{mix}}}
\newcommand{\Var}{\operatorname{Var}}
\newcommand{\diam}{\operatorname{diam}}

\definecolor{hred}{HTML}{A3132F}
\definecolor{hblue}{HTML}{164C9A}
\definecolor{lred}{HTML}{F3C3CC}
\definecolor{lblue}{HTML}{C3D7F0}
\definecolor{teal}{HTML}{0B7F86}
\definecolor{amber}{HTML}{B85C00}
\definecolor{gridgray}{HTML}{C8CDD2}
\definecolor{cellgray}{HTML}{BFC3C9}
\definecolor{cellblue}{HTML}{7FB8CC}
\definecolor{cellgold}{HTML}{E3A63E}

\tikzset{
  mid arrow/.style={postaction={decorate},decoration={markings,
      mark=at position #1 with {\arrow{Stealth[length=5pt,width=4pt]}}}},
  mid arrow/.default=0.62,
  gridline/.style={gridgray,line width=0.5pt},
  medarc/.style={teal,line width=0.7pt,mid arrow=0.55},
  wtwo/.style={circle,fill=amber,draw=amber,inner sep=0pt,minimum size=4.2pt},
  wone/.style={circle,fill=white,draw=teal,line width=0.5pt,inner sep=0pt,minimum size=4.2pt},
  hue/.style={circle,draw=none,inner sep=0pt,minimum size=13pt,text=white,font=\scriptsize\sffamily},
  shade/.style={circle,draw=black!35,line width=0.3pt,inner sep=0pt,minimum size=15pt,font=\tiny\sffamily},
  pvtx/.style={circle,fill=black!80,inner sep=0pt,minimum size=3.2pt},
}

\title{Efficient algorithms for uniformly sampling four- and \\ five-colorings of grid regions} 

\author{Jacob Platnick$^*$ \ and \ Dana Randall\thanks{School of Computer Science, Georgia Institute of Technology, Atlanta, GA 30332. Email: {\tt jplatnick3@gatech.edu, randall@cc.gatech.edu.}}}

\date{}

\begin{document}
\maketitle

% ======================================================================
\begin{abstract}
A proper $q$-coloring of a graph assigns one of $q$ colors to each vertex so that adjacent vertices receive different colors, corresponding to the zero-temperature antiferromagnetic Potts model from statistical physics. On rectangular square-grid regions with free boundary, Glauber dynamics is known to mix rapidly for three colors and for every $q\ge 6$, while the cases $q=4$ and $q=5$ have remained open for about two decades. We give efficient algorithms for both, using a different method for each.

For four colors we give a fully polynomial randomized approximation scheme (FPRAS) for the number of proper four-colorings of an arbitrary bipartite planar graph, and an efficient sampling algorithm. The class contains every finite subgraph of $\Z^2$: rectangular grids and grid domains with or without holes. On a simple graph with $N$ vertices, the counting algorithm achieves relative error $\delta$ in time $O(N^{13}\delta^{-2}\log^2(2N/\delta))$. %It also admits an implementation whose bit complexity is polynomial in the input length and $1/\delta$.} 
The algorithm does not run a Markov chain on colorings directly. We use Rys' spin representation to show that the number of four-colorings equals $2^{N-M}$ times the partition function of a six-vertex model with weights $(2,1,1)$ on the medial graph, where $M$ is the number of edges, and we approximate this partition function with the recent FPRAS of Chen and Liu for counting the common bases of two matroids.

For five colors we show that Glauber dynamics itself is rapidly mixing on rectangular grid regions. On an $h\times w$ rectangle with free boundary, a heat bath that resamples two adjacent rows at a time mixes in $O(s\log(n/\varepsilon))$ updates, where $n=hw$ and $s$ is the shorter side, and Glauber dynamics has inverse spectral gap $O(ns^3)$. The analysis rests on a coupling estimate for two-row strips that was checked by an integer verifier.
\end{abstract}

% ======================================================================
\section{Introduction}\label{sec:intro}

A proper $q$-coloring of a graph $G=(V,E)$ assigns one of $q$ colors to each vertex so that adjacent vertices receive different colors. Uniformly random proper colorings are the zero-temperature limit of the antiferromagnetic $q$-state Potts model from statistical physics~\cite{Sok05}, and generating them efficiently is a basic problem both there and in theoretical computer science. Physicists sample colorings to estimate correlations and other thermodynamic quantities. In computer science, sampling and counting are closely related: for self-reducible problems such as colorings, an efficient almost uniform sampler yields a fully polynomial randomized approximation scheme (FPRAS) for the number of solutions, and conversely, by the general reduction of Jerrum, Valiant and Vazirani~\cite{JVV86}. Exact counting is out of reach in general, since the number of proper $q$-colorings is an evaluation of the chromatic polynomial and is \#P-hard to compute for every fixed $q\ge 3$~\cite{JVW90}.

The standard tool for sampling and approximate counting is Markov chain Monte Carlo. One designs a Markov chain whose states are the proper colorings, whose moves make small local changes, and whose stationary distribution is uniform. After sufficiently many steps the current state is close to a uniform sample. The question is how many steps suffice. The mixing time $\tmix(\varepsilon)$ is the number of steps needed to come within $\varepsilon$ of stationarity in total variation distance, from the worst starting state. A chain is rapidly mixing if $\tmix(\varepsilon)$ is bounded by a polynomial in the size of $G$ and $\log(1/\varepsilon)$.

The simplest chain is single-site Glauber dynamics: choose a vertex and a color uniformly at random, and recolor the vertex if the result is proper. On graphs of maximum degree $\Delta$ it is rapidly mixing whenever $q>2\Delta$~\cite{Jer95,SS97}, and path coupling~\cite{BD97} gives a particularly simple proof. Bubley, Dyer, Greenhill and Jerrum showed, with computer assistance, that the factor $2$ is not necessary for graphs of small degree~\cite{BDG98,BDGJ99}. Vigoda lowered the general threshold to $q>\frac{11}{6}\Delta$~\cite{Vig00}, and Carlson and Vigoda lowered it to $q\ge 1.809\Delta$ for $\Delta\ge 125$~\cite{CV25}. These bounds hold for every graph. On particular lattices one can hope to do much better, and the square lattice $\Z^2$, where $\Delta=4$, has long served as the main test case.

Three-colorings of $\Z^2$ are well understood because they have special structure. Up to a global choice, they are in bijection with configurations of square ice, whose entropy Lieb computed exactly~\cite{Lie67}. Equivalently, a three-coloring of a simply connected region lifts to a height function: an integer-valued function whose values at adjacent vertices differ by exactly one and which reduces to the coloring modulo~$3$. Height functions are partially ordered, and this order makes monotone coupling arguments available. Luby, Randall and Sinclair~\cite{LRS01} exploited this structure. They represented three-colorings of a region with fixed boundary colors as families of nonintersecting lattice paths, and gave the first rigorous polynomial bounds for Markov chains that sample them; the same framework covers domino and lozenge tilings. Randall and Tetali~\cite{RT00} then used the comparison method to show that Glauber dynamics itself mixes in polynomial time in these settings, and Goldberg, Martin and Paterson~\cite{GMP04} proved that Glauber dynamics is rapidly mixing on rectangles with free boundary. Mihail and Winkler~\cite{MW96} gave a polynomial time  algorithm for sampling Eulerian orientations on Eulerian graphs, which can used to sample $3$-colorings on simply connected induced subgraphs of rectangular grids by sampling Eulerian orientations on a subdivided dual multigraph.

For $q\ge 6$ the general results above give a starting point. On the square lattice the condition $q>2\Delta$ means $q\ge 9$~\cite{Jer95,SS97}, and Vigoda's bound gives $q\ge 8$~\cite{Vig00}. Lattice-specific arguments go further. Goldberg, Martin and Paterson~\cite{GMP05} extended recursive couplings from trees to lattices, and their results include seven colors on triangle-free $4$-regular graphs. Achlioptas, Molloy, Moore and Van Bussel~\cite{AMMV04}, and Goldberg, Jalsenius, Martin and Paterson~\cite{GJMP06}, proved strong spatial mixing for six colors on the square lattice. By the equivalence between spatial and temporal mixing~\cite{Mar99,DSVW04}, strong spatial mixing implies rapid mixing for block dynamics and for Glauber dynamics.

\auditrevision{On rectangular square-grid regions with free boundary, rapid mixing of Glauber dynamics is thus known for three colors and for six or more colors.} Achlioptas, Molloy, Moore and Van Bussel~\cite{AMMV04} noted four and five colors as the only remaining cases. Both have remained open for about two decades.

Five colors lie at the threshold $q=\Delta+1$ of the square lattice and four colors at $q=\Delta$, and neither case is covered by the height-function arguments that work for three colors. With five colors, consider a vertex whose three other neighbors carry three distinct colors, none of them $s$ or $t$. If its fourth neighbor changes from $s$ to $t$, the vertex is forced to change from $t$ to $s$. So a single change can propagate with probability one, and no contraction can come from looking at one site at a time. With four colors local moves are even more constrained: a vertex whose four neighbors carry the three other colors cannot change at all under single-site dynamics. On the torus $\Z_n^2$ with $4\mid n$, the coloring $(x,y)\mapsto (x+2y)\bmod 4$ has this property at every vertex, since the neighbors of a vertex of color $c$ have colors $c\pm1$ and $c+2$. This coloring is frozen, so Glauber dynamics is not even irreducible there. On rectangular grids with free boundary we do not know whether Glauber dynamics for four colors is rapidly mixing.

In higher dimensions local dynamics can be exponentially slow. For three colors, Galvin and Randall~\cite{GR07} and Galvin, Kahn, Randall and Sorkin~\cite{GKRS15} showed that uniformly random colorings of high-dimensional lattice regions exhibit long-range order: a typical coloring follows one of six ordered patterns, in which one sublattice is predominantly a single color and the other sublattice uses the remaining two (see also Peled~\cite{Pel17}). Moving from one pattern to another requires passing through colorings that contain an interface between ordered regions, and such colorings carry exponentially small probability. This bottleneck forces Glauber dynamics, and indeed any sufficiently local Markov chain, to take exponential time~\cite{GR07,GKRS15}. Peled and Spinka~\cite{PS23} proved long-range order for every $q\ge 3$, confirming a conjecture of Galvin, Kahn, Randall and Sorkin~\cite[Conjecture~1.3]{GKRS15}. When $d\ge Cq^{10}\log^2 q$, a uniformly random proper $q$-coloring follows, with high probability at each vertex, one of the dominant patterns in which $\lfloor q/2\rfloor$ colors occupy one sublattice and the remaining $\lceil q/2\rceil$ colors the other. For four colors there are $\binom{4}{2}=6$ such patterns, with two colors on the even sites and two on the odd sites, and for five colors there are $2\binom{5}{2}=20$. As for three colors, this long-range order implies that local Markov chains for four- and five-colorings require exponential time on high-dimensional lattice regions. Efficient algorithms must therefore exploit structure that is absent in high dimensions. For four colors that structure is planarity, and for five colors it is the two-dimensional geometry of the square lattice.

% ----------------------------------------------------------------------
\subsection{Our results}

We give efficient algorithms for both remaining cases, with a different method for each.

All graphs are finite. For a graph $G$, write $\cC_4(G)$ for its set of proper four-colorings, $Z_4(G)=|\cC_4(G)|$, and $U_G$ for the uniform distribution on $\cC_4(G)$. No boundary conditions are imposed. Our first result is an FPRAS for $Z_4$ on bipartite planar graphs.

\begin{theorem}[Four colors]\label{four:thm:main}
There is a randomized algorithm that, given a bipartite planar graph $G$ and a rational number $0<\delta<1$, returns $\widehat Z$ such that
\[
  \Pr\bigl[(1-\delta)Z_4(G)\le \widehat Z\le(1+\delta)Z_4(G)\bigr]\ge \tfrac34 .
\]
For a simple graph with $N\ge 1$ vertices, the algorithm uses
\[
  O\bigl(N^{13}\,\delta^{-2}\log^2(2N/\delta)\bigr)
\]
arithmetic operations. %It admits an implementation whose running time in bit operations is polynomial in the input length and $1/\delta$.}
\end{theorem}

The input need not come with an embedding; one is computed in linear time. Parallel edges do not change the set of proper colorings and are deleted first. As usual, the success probability can be raised to $1-\gamma$ by taking the median of $O(\log(1/\gamma))$ independent runs. By the reduction of Jerrum, Valiant and Vazirani, the counting algorithm also gives a sampler.

\begin{corollary}\label{four:cor:sampling}
There is a randomized algorithm that, given a bipartite planar graph $G$ and $0<\eta<1$, outputs a proper four-coloring of $G$ whose distribution is within total variation distance $\eta$ of $U_G$, in time polynomial in the size of $G$ and $\log(1/\eta)$.
\end{corollary}

These results cover every rectangular grid $P_a\,\square\,P_b$ and every finite subgraph of $\Z^2$, induced or not, simply connected or with holes. This is exactly the setting in which rapid mixing of Glauber dynamics for four colors has been open. Planarity is essential to the method, in line with the high-dimensional picture above.

For five colors we show that Glauber dynamics itself is rapidly mixing on rectangular grids. Let $P_h\,\square\,P_w$ be the grid with $h$ rows and $w$ columns, $n=hw$ sites and shorter side $s=\min(h,w)$, with no colors prescribed outside it. The \emph{two-row chain} chooses two adjacent rows uniformly at random and replaces their colors by a uniformly random proper coloring compatible with the rows just above and just below. For five colors, $0<\varepsilon<1/2$ throughout.

\begin{theorem}[Five colors: the two-row chain]\label{five:thm:tworow}
On the $h\times w$ rectangle with free boundary and $h\ge2$, the ideal two-row chain is reversible for the uniform distribution and has inverse spectral gap at most $(h-1)/(1-c)$, where $c=999970/10^6$. It has an implementation with boundedly many fair bits and polynomial work per update whose mixing time satisfies
\[
  \tmix(\varepsilon)\le\bigl\lceil 10^5(h-1)\log(n/\varepsilon)\bigr\rceil.
\]
Orienting the rectangle so that $h=s$ gives at most $\lceil 10^5(s-1)\log(n/\varepsilon)\rceil$ updates. For a single row a whole-path update gives $O(1+\log(1/\varepsilon))$.
\end{theorem}

\begin{theorem}[Five colors: Glauber dynamics]\label{five:thm:glauber}
On the $h\times w$ rectangle with free boundary, Glauber dynamics for five colors has inverse spectral gap $O(ns^3)$, and hence
\[
  \tmix(\varepsilon)=O\bigl(ns^3(n+\log(1/\varepsilon))\bigr).
\]
On a path it mixes in $O(n\log(n/\varepsilon))$ steps.
\end{theorem}

% ----------------------------------------------------------------------
\subsection{Techniques}

The two methods are unrelated. For four colors we do not analyze any Markov chain on colorings. Instead we reduce approximate counting of four-colorings to approximate counting in a six-vertex model, for which an FPRAS follows from the recent FPRAS of Chen and Liu for counting the common bases of two matroids~\cite{CL26}, a long-standing open problem. It will be convenient to refer to the four colors with \emph{hues and shades}. Group the four colors into two hues, red and blue, each with a dark and a light shade. A coloring is then a hue pattern together with a shade at each vertex. Within a maximal connected region of one hue, the two shades must alternate along every edge, and since $G$ is bipartite there are exactly two ways to do this. Different regions are independent. So a hue pattern with $k$ regions extends to exactly $2^k$ proper colorings (Lemma~\ref{four:lem:shades}).

Give weight $2$ to each medial vertex whose primal edge joins two different hues. An Euler-formula computation shows that every hue pattern with $k$ regions receives total weight $2^{M-N}2^{k}$. Hence
\[
  Z_4(G)=2^{N-M}\,Y_6(\Med G),
\]
where $Y_6$ is the partition function of the six-vertex model with local weights $(2,1,1)$ (Lemma~\ref{four:lem:identity}). These weights lie on the boundary of the region covered by the six-vertex FPRAS of Chen and Liu~\cite{CL26}. The remaining hypothesis of that result, that the medial graph has no loops, is achieved by first pruning vertices of degree at most one, which multiplies the count by a known factor. Sampling follows from the self-reducibility framework of Jerrum, Valiant and Vazirani~\cite{JVV86}: sample a weighted orientation, decode the hue pattern, and choose the shades in each hue region with one fair coin.

For five colors we analyze Markov chains directly. Both theorems for rectangles rest on one estimate, Lemma~\ref{five:lem:strip}: if a single exterior color next to a two-row strip changes, the two conditional laws of the strip can be coupled with at most $c<1$ expected disagreements, uniformly in the length of the strip. The uniform coloring of a strip is a Markov chain along its columns, each of which takes one of only $20$ values, and the estimate is established by a transfer recurrence and a finite certificate checked in integer arithmetic {\color{changes}(Appendix~\ref{app:certificate})}. With the estimate in hand, path coupling for the two-row chain is short: two copies choose the same strip, disagreements inside the strip are erased, and each disagreement in an adjacent row creates at most $c$ new ones. Glauber dynamics is then handled by the comparison method.

Note that even if some block dynamics such as the two-row chain were shown to mix rapidly for four colors, rapid mixing of Glauber dynamics would not follow automatically. With five colors, each site of a two-row strip keeps at least three available colors when the exterior and the other row are fixed, one more than its neighbors in its own row, so every strip update can be simulated by single-site moves inside the strip. With four colors only two may remain, and single-site moves can get stuck: for example, if the rows of a rectangle repeat the patterns $abab\cdots$, $cdcd\cdots$, $baba\cdots$, $dcdc\cdots$, then no site outside the top and bottom rows can be recolored, although a strip can still change as a block.  It follows that the comparison paths on four colorings might need routes that change sites far outside the strip itself.

The four-color reduction builds on the correspondence of Baxter, Kelland and Wu~\cite{BKW76} between the Potts model on a planar graph and an ice-type model on its medial graph, which we use in the form of Rys' spin representation~\cite{Rys63} given by Lis~\cite{Lis22}. For six colors, Achlioptas, Molloy, Moore and Van Bussel coupled $2\times3$ heat baths under a single exterior disagreement and compared the resulting block dynamics with Glauber dynamics~\cite{AMMV04}; for five colors we need two-row strips of arbitrary length. Our strip recurrence extends the classical transfer matrices for square-lattice chromatic polynomials~\cite{SS01} to transport distances, in the spirit of the block and recursive couplings of~\cite{DSVW04,GMP05}, and the comparison with local chains follows Diaconis and Saloff-Coste~\cite{DSC93} and Randall and Tetali~\cite{RT00}.

%Section~\ref{four:sec:part} treats four colors and Section~\ref{five:sec:rect-part} treats five colors. The treatments of four and five colors are independent and can be read in either order. Section~\ref{sec:remarks} contains concluding remarks.

% ======================================================================
\section{Four-colorings of bipartite planar graphs}\label{four:sec:part}

This section proves Theorem~\ref{four:thm:main} and Corollary~\ref{four:cor:sampling}.

\subsection{Preliminaries}\label{four:sec:prelim}

Let $f$ be a function from problem instances to nonnegative numbers. An FPRAS for $f$ is a randomized algorithm that, given an instance $x$ and $0<\delta<1$, outputs $\widehat Z$ with $\Pr[(1-\delta)f(x)\le\widehat Z\le(1+\delta)f(x)]\ge 3/4$, in time polynomial in the size of $x$ and $1/\delta$. The total variation distance between distributions $\mu,\nu$ on a finite set $\Omega$ is $\|\mu-\nu\|_\TV=\frac12\sum_{x\in\Omega}|\mu(x)-\nu(x)|$. Applying the same map to two random variables cannot increase the total variation distance between their laws~\cite{LP17}.

Until Section~\ref{four:sec:counting}, $G=(V,E)$ is a nonempty finite connected bipartite plane graph, given with its embedding; a rotation system suffices. Parallel edges are allowed. We write $N=|V|$, $M=|E|$ and $F(G)$ for the set of faces, and fix one face $\finf$, the outer face. Let $\eps:V\to\{0,1\}$ be a proper two-coloring of $G$. We call $v$ white if $\eps(v)=0$ and black if $\eps(v)=1$. A four-coloring is a map $c:V\to\{0,1\}^2$, and it is proper if adjacent vertices receive different colors.

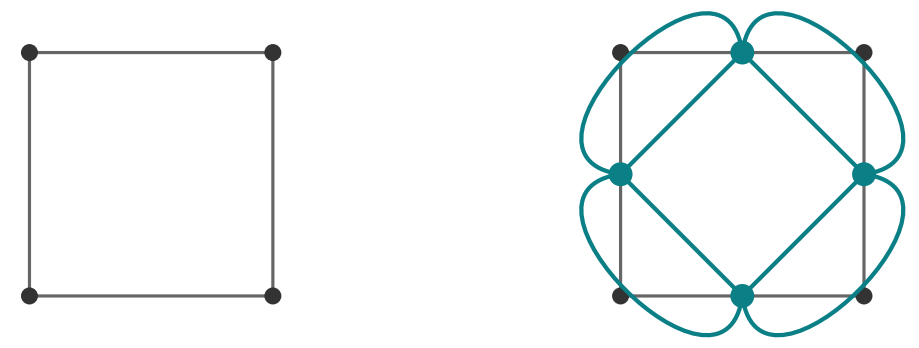
\begin{figure}[t]
\centering
\begin{tikzpicture}[scale=1.15]
  \begin{scope}
    \draw[black!60,line width=0.6pt] (0,0) rectangle (1.4,1.4);
    \foreach \p in {(0,0),(1.4,0),(1.4,1.4),(0,1.4)} \node[pvtx] at \p {};
    \node[font=\small,text=black!70] at (0.7,-0.9) {Plane graph $G$};
  \end{scope}
  \begin{scope}[xshift=3.4cm]
    \draw[black!60,line width=0.6pt] (0,0) rectangle (1.4,1.4);
    \foreach \p in {(0,0),(1.4,0),(1.4,1.4),(0,1.4)} \node[pvtx] at \p {};
    \coordinate (B) at (0.7,0);   \coordinate (R) at (1.4,0.7);
    \coordinate (T) at (0.7,1.4); \coordinate (L) at (0,0.7);
    \draw[teal,line width=0.8pt] (B)--(R)--(T)--(L)--cycle;
    \draw[teal,line width=0.8pt] (T) .. controls +(-0.1,0.75) and +(-0.75,0.1) .. (L);
    \draw[teal,line width=0.8pt] (T) .. controls +(0.1,0.75) and +(0.75,0.1) .. (R);
    \draw[teal,line width=0.8pt] (B) .. controls +(-0.1,-0.75) and +(-0.75,-0.1) .. (L);
    \draw[teal,line width=0.8pt] (B) .. controls +(0.1,-0.75) and +(0.75,-0.1) .. (R);
    \foreach \p in {B,R,T,L} \node[circle,fill=teal,inner sep=0pt,minimum size=4.5pt] at (\p) {};
    \node[font=\small,text=black!70] at (0.7,-0.9) {Medial graph over $G$};
  \end{scope}
\end{tikzpicture}
\caption{The four edges of a square give four medial vertices. At each corner of the square, one interior arc and one exterior arc join the same two midpoints. Both belong to the medial graph (teal).}
\label{four:fig:medial}
\end{figure}

A \emph{corner} of $G$ is an incidence of a vertex $v$ with a face $f$ between two consecutive edges $e,e'$ in the rotation at $v$. A face that meets $v$ several times gives several corners, and a vertex of degree one has a single corner, between its edge and itself. There are $\sum_v \deg v=2M$ corners. For a connected plane graph $G$ with at least one edge, its \emph{medial graph}~\cite{Lis22} $H=\Med(G)$ has one vertex $v_e$ at the midpoint of each edge $e$. Each corner of $G$ joins its two incident edge midpoints by a medial edge, including the corners of the outer face (Figure~\ref{four:fig:medial}). Thus $H$ is a $4$-regular plane multigraph with $M$ vertices and $2M$ edges. It has a loop exactly when $G$ has a vertex of degree one.

A \emph{six-vertex configuration} on a $4$-regular graph $H$ is an Eulerian orientation of $H$: every vertex has exactly two incoming edges. At a vertex $x$ the six possible incoming pairs of half-edges split into three complementary pairs $\{P,P^c\}$, and a weight function at $x$ assigns a nonnegative weight to each of the three classes. The weight of a configuration is the product over all vertices of the weight of the incoming pair, and the partition function $Y_6(H)$ is the sum of these weights over all six-vertex configurations. The weights may differ from vertex to vertex.

% ======================================================================
\subsection{Hues and shades}\label{four:sec:identity}

From here until Section~\ref{four:sec:counting}, $G$ is as in Section~\ref{four:sec:prelim} and $M\ge1$.

Each vertex has a \emph{hue}, red or blue, and a \emph{shade}, dark or light. We write its color as $c(v)=(r(v),s(v))$, encoding red and blue by $r=0,1$ and dark and light by $s=0,1$. A \emph{hue region} is a maximal connected set of vertices of one hue. For a hue assignment $r:V\to\{0,1\}$, let
\[
  S=S(r)=\{uv\in E:\ r(u)=r(v)\},\qquad k(S)=\#\{\text{components of }(V,S)\}.
\]
The components counted by $k(S)$ are exactly the hue regions, including singletons.

\begin{lemma}[Shades within hue regions]\label{four:lem:shades}
Each hue assignment $r$ has exactly $2^{k(S(r))}$ proper shade assignments. Hence
\begin{equation}\label{four:eq:hue-sum}
  Z_4(G)=\sum_{r:V\to\{0,1\}}2^{k(S(r))}.
\end{equation}
\end{lemma}

\begin{proof}
Within a red region, dark red and light red must alternate along every edge, and the same holds for the two blue shades within a blue region. On a connected hue region, the shade at one vertex determines all others by alternation, and the two choices are exactly the restrictions of $s=\eps$ and $s=1-\eps$ to that region; both are consistent because $G$ is bipartite. Across a red--blue edge either shade is allowed at either endpoint, so these choices are independent across hue regions. Summing over $r$ gives~\eqref{four:eq:hue-sum}.
\end{proof}

\subsection{The spin--arrow correspondence}

Rys' mapping~\cite{Rys63} encodes the vertex hues together with an auxiliary face-spin assignment $h:F(G)\to\{0,1\}$. A hue assignment and a face-spin assignment are \emph{compatible} if the two faces on either side of every red--blue edge have equal spins. Along a same-hue edge the face spins may agree or differ.

\begin{lemma}[Spin--arrow correspondence~{\cite[Lemma~3.7]{Lis22}}]\label{four:lem:spin-arrow}
Compatible pairs $(r,h)$ with $h(\finf)=0$ are in bijection with the Eulerian orientations of the medial graph $\Med(G)$.
\end{lemma}

\begin{proof}
Given a compatible pair $(r,h)$, at the corner between a vertex $v$ and a face $f$, orient the medial edge clockwise around $v$ if $r(v)=h(f)$, and counterclockwise otherwise.

To see that this orientation is Eulerian exactly when the pair is compatible, consider an edge $e=uv$, drawn horizontally with $u$ on the left, and let $f$ and $g$ be the faces above and below it ($f=g$ if $e$ is a bridge). The four medial edges at $v_e$ come from the corners $(u,f)$, $(v,f)$, $(u,g)$ and $(v,g)$, which we call NW, NE, SW and SE. If both endpoints are red and both spins are zero, every corner is oriented clockwise, and the NW and SE arrows enter $v_e$ while the NE and SW arrows leave. Changing the hue of an endpoint reverses its two arrows, and changing the spin of a face reverses the two arrows on that side. Writing $\alpha=r(u)$, $\beta=r(v)$, $s=h(f)$, $t=h(g)$ and $[\cdot]$ for the indicator of an event, the number of arrows entering $v_e$ is therefore
\[
  [\alpha=s]+[\beta\neq s]+[\alpha\neq t]+[\beta=t].
\]
If $\alpha=\beta$ this equals $2$ for all $s,t$. If $\alpha\ne\beta$ it equals $2\bigl([\alpha=s]+[\alpha\ne t]\bigr)$, which is $2$ if $s=t$ and $0$ or $4$ if $s\ne t$. So the orientation is Eulerian at $v_e$ exactly when $(r,h)$ is compatible at $e$. Up to reversing all arrows, the three compatible patterns are those of Figure~\ref{four:fig:types}.

For the inverse map, form the radial graph of $G$, with a node for each vertex and each face of $G$ and one edge for each corner, joining its vertex and its face. Given an Eulerian orientation, label a corner $0$ if its arrow is clockwise around its vertex and $1$ otherwise. Each face of the radial graph surrounds a single medial vertex, and its four labels have even sum: the given orientation and the all-clockwise orientation each have two incoming arrows there, so their incoming pairs differ in an even number of positions. Here repeated incidences are counted with multiplicity, including at a bridge. The facial walks generate the binary cycle space of the radial graph on the sphere~\cite{Die17} (summing the faces inside a simple cycle cancels each interior edge twice), so every closed walk has even label sum. Assign $\finf$ the value $0$, and assign every other radial node the sum modulo two of the labels along any path from $\finf$. The radial graph is connected, and the closed-walk identity makes this assignment independent of the path. The resulting hues and spins satisfy $r(v)\oplus h(f)=\text{label of the corner }(v,f)$, where $\oplus$ is addition modulo two, so they reconstruct every arrow. By the computation above the pair is compatible, and connectivity gives uniqueness once $h(\finf)$ is fixed.
\end{proof}

\subsection{The weights and the identity}

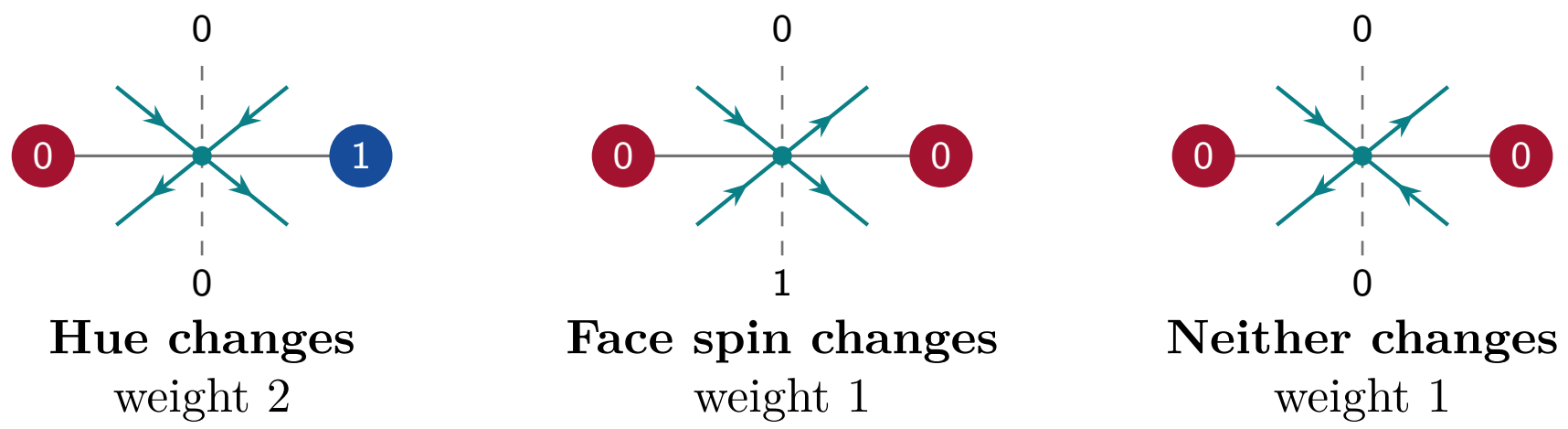
\begin{figure}[t]
\centering
\begin{tikzpicture}[scale=1.0]
  % one panel: #1 = x shift, hues and spins, and in/out pattern
  \newcommand{\edgepanel}[9]{%
    \begin{scope}[xshift=#1]
      \draw[black!55,line width=0.6pt] (-1.15,0)--(1.15,0);
      \draw[black!55,dashed,line width=0.5pt] (0,-0.72)--(0,0.72);
      \node[hue,fill=#2] at (-1.15,0) {#3};
      \node[hue,fill=#4] at (1.15,0) {#5};
      \node[font=\scriptsize\sffamily] at (0,0.92) {#6};
      \node[font=\scriptsize\sffamily] at (0,-0.92) {#7};
      #8
      \node[circle,fill=teal,inner sep=0pt,minimum size=4pt] at (0,0) {};
      \node[align=center,font=\small] at (0,-1.55) {#9};
    \end{scope}}
  \tikzset{arr/.style={teal,line width=0.8pt,mid arrow=0.6}}
  \edgepanel{0cm}{hred}{0}{hblue}{1}{0}{0}%
    {\draw[arr] (-0.62,0.5)--(0,0); \draw[arr] (0.62,0.5)--(0,0);
     \draw[arr] (0,0)--(-0.62,-0.5); \draw[arr] (0,0)--(0.62,-0.5);}%
    {\textbf{Hue changes}\\ weight $2$}
  \edgepanel{4.2cm}{hred}{0}{hred}{0}{0}{1}%
    {\draw[arr] (-0.62,0.5)--(0,0); \draw[arr] (0,0)--(0.62,0.5);
     \draw[arr] (-0.62,-0.5)--(0,0); \draw[arr] (0,0)--(0.62,-0.5);}%
    {\textbf{Face spin changes}\\ weight $1$}
  \edgepanel{8.4cm}{hred}{0}{hred}{0}{0}{0}%
    {\draw[arr] (-0.62,0.5)--(0,0); \draw[arr] (0,0)--(0.62,0.5);
     \draw[arr] (0,0)--(-0.62,-0.5); \draw[arr] (0.62,-0.5)--(0,0);}%
    {\textbf{Neither changes}\\ weight $1$}
\end{tikzpicture}
\caption{One primal edge, drawn horizontally. The circles show the endpoint hues (red $0$, blue $1$) and the numbers above and below give the face spins. Reversing all four arrows gives the other orientation of each type. The weight-two pair can enter from either side.}
\label{four:fig:types}
\end{figure}

Give a medial vertex weight $2$ if both of its incoming arrows lie on the same side of the primal edge, and weight $1$ otherwise. In the notation of the proof of Lemma~\ref{four:lem:spin-arrow}, the incoming pair is $\{\mathrm{NW},\mathrm{NE}\}$ or $\{\mathrm{SW},\mathrm{SE}\}$ when $\alpha\ne\beta$ and $s=t$, and it contains one arrow from each side when $\alpha=\beta$. So a medial vertex has weight $2$ precisely when the hues at the ends of its primal edge differ. The three complementary classes of incoming pairs receive weights $2,1,1$ (Figure~\ref{four:fig:types}). In the notation of Lis~\cite{Lis22} we assign $(a,b)=(1,2)$ directly to the combinatorial types; his spin law is stated for $a,b\le1$.

\begin{lemma}\label{four:lem:identity}
With these local weights, $Z_4(G)=2^{N-M}\,Y_6(\Med G)$.
\end{lemma}

\begin{proof}
Fix the hues $r$, and draw the spanning subgraph $(V,S)$ by erasing every red--blue edge. Erasing an edge merges its two incident faces if they are distinct. Compatibility requires the same spin on both sides of every erased edge, so a compatible spin assignment is exactly a choice of one spin per face of $(V,S)$, with the spin of the outer face fixed to zero.

Write $q$ for the number of faces of $(V,S)$. Euler's formula for a plane graph with $k(S)$ components gives $q-1=|S|-N+k(S)$~\cite{Die17}. By Lemma~\ref{four:lem:spin-arrow}, each of the $2^{q-1}$ spin choices gives one Eulerian orientation, and each such orientation has weight $2^{M-|S|}$, one factor of two per red--blue edge. Thus the total weight of the orientations with hue assignment $r$ is
\[
  2^{q-1}\,2^{M-|S|}=2^{M-N}\,2^{k(S)}.
\]
Summing over $r$ and applying~\eqref{four:eq:hue-sum} proves the identity.
\end{proof}

\begin{figure}[p]
\centering
\begin{tikzpicture}[scale=0.62]
  % ------------------------------------------------ hue regions and S
  \begin{scope}
    \foreach \i/\row in {0/{0,0,1,1,1},1/{0,1,1,1,1},2/{0,1,0,0,0},3/{1,1,0,1,1}}{
      \foreach \h [count=\j from 0] in \row { \expandafter\xdef\csname H\i-\j\endcsname{\h} } }
    % edges
    \foreach \i in {0,...,3}{\foreach \j in {0,...,3}{
      \pgfmathtruncatemacro{\jj}{\j+1}
      \edef\a{\csname H\i-\j\endcsname}\edef\b{\csname H\i-\jj\endcsname}
      \ifx\a\b \draw[teal,line width=1.6pt] (\j,-\i)--(\jj,-\i); \else \draw[gridline] (\j,-\i)--(\jj,-\i); \fi }}
    \foreach \i in {0,...,2}{\foreach \j in {0,...,4}{
      \pgfmathtruncatemacro{\ii}{\i+1}
      \edef\a{\csname H\i-\j\endcsname}\edef\b{\csname H\ii-\j\endcsname}
      \ifx\a\b \draw[teal,line width=1.6pt] (\j,-\i)--(\j,-\ii); \else \draw[gridline] (\j,-\i)--(\j,-\ii); \fi }}
    \foreach \i in {0,...,3}{\foreach \j in {0,...,4}{
      \edef\a{\csname H\i-\j\endcsname}
      \if0\a \node[hue,fill=hred] at (\j,-\i) {0}; \else \node[hue,fill=hblue] at (\j,-\i) {1}; \fi }}
    \node[font=\small\bfseries] at (2,-4.1) {Hue regions and $S$};
  \end{scope}
  % ------------------------------------------------ compatible face spin
  \begin{scope}[xshift=7.2cm]
    \fill[teal!12] (3,0) rectangle (4,-1);
    \foreach \i in {0,...,3} \draw[gridline] (0,-\i)--(4,-\i);
    \foreach \j in {0,...,4} \draw[gridline] (\j,0)--(\j,-3);
    \foreach \i in {0,...,3}{\foreach \j in {0,...,4} \fill[black!60] (\j,-\i) circle (1.6pt);}
    \foreach \i in {0,1,2}{\foreach \j in {0,1,2,3}{
      \ifnum\i=0 \ifnum\j=3 \node[font=\small\sffamily] at (\j+0.5,-\i-0.5) {1}; \else \node[font=\small\sffamily] at (\j+0.5,-\i-0.5) {0}; \fi
      \else \node[font=\small\sffamily] at (\j+0.5,-\i-0.5) {0}; \fi }}
    \draw[amber,line width=1.6pt] (3,0) rectangle (4,-1);
    \node[font=\small] at (2,0.6) {$h(\finf)=0$};
    \node[font=\small\bfseries] at (2,-4.1) {A compatible face spin};
  \end{scope}
  % ------------------------------------------------ hues and shades
  \begin{scope}[xshift=14.4cm]
    \foreach \i in {0,...,3} \draw[gridline] (0,-\i)--(4,-\i);
    \foreach \j in {0,...,4} \draw[gridline] (\j,0)--(\j,-3);
    \foreach \i/\row in {0/{00,01,11,10,11},1/{01,11,10,11,10},2/{00,10,00,01,00},3/{10,11,01,10,11}}{
      \foreach \cl [count=\j from 0] in \row {
        \ifnum\cl=0  \node[shade,fill=hred,text=white] at (\j,-\i) {00}; \fi
        \ifnum\cl=1  \node[shade,fill=lred] at (\j,-\i) {01}; \fi
        \ifnum\cl=10 \node[shade,fill=hblue,text=white] at (\j,-\i) {10}; \fi
        \ifnum\cl=11 \node[shade,fill=lblue] at (\j,-\i) {11}; \fi } }
    \node[font=\small\bfseries] at (2,-4.1) {Hues and shades};
  \end{scope}
\end{tikzpicture}

\vspace{0.6cm}
\begin{tikzpicture}[scale=1.25]
% medial orientation, generated
\draw[gridline] (2,-1)--(1,-1);
\draw[gridline] (4,0)--(3,0);
\draw[gridline] (2,-2)--(1,-2);
\draw[gridline] (3,-1)--(3,0);
\draw[gridline] (0,-3)--(0,-2);
\draw[gridline] (2,-3)--(1,-3);
\draw[gridline] (2,-3)--(2,-2);
\draw[gridline] (4,-2)--(3,-2);
\draw[gridline] (4,-2)--(4,-3);
\draw[gridline] (3,-3)--(4,-3);
\draw[gridline] (2,-1)--(3,-1);
\draw[gridline] (2,-2)--(3,-2);
\draw[gridline] (3,-3)--(3,-2);
\draw[gridline] (0,-1)--(0,-2);
\draw[gridline] (1,-2)--(1,-1);
\draw[gridline] (2,-1)--(2,-2);
\draw[gridline] (1,0)--(0,0);
\draw[gridline] (3,-1)--(3,-2);
\draw[gridline] (1,0)--(1,-1);
\draw[gridline] (4,-1)--(4,-2);
\draw[gridline] (1,-3)--(1,-2);
\draw[gridline] (2,0)--(3,0);
\draw[gridline] (1,0)--(2,0);
\draw[gridline] (2,-3)--(3,-3);
\draw[gridline] (0,-3)--(1,-3);
\draw[gridline] (2,-1)--(2,0);
\draw[gridline] (0,-1)--(1,-1);
\draw[gridline] (0,-2)--(1,-2);
\draw[gridline] (0,-1)--(0,0);
\draw[gridline] (4,-1)--(4,0);
\draw[gridline] (4,-1)--(3,-1);
\draw[medarc] (0.5,0) -- (0,-0.5);
\draw[medarc] (0,-0.5) .. controls (-1,0) and (0,1) .. (0.5,0);
\draw[medarc] (1.5,0) -- (1,-0.5);
\draw[medarc] (1,-0.5) -- (0.5,0);
\draw[medarc] (0.5,0) .. controls (1,0.9) .. (1.5,0);
\draw[medarc] (2,-0.5) -- (2.5,0);
\draw[medarc] (1.5,0) -- (2,-0.5);
\draw[medarc] (2.5,0) .. controls (2,0.9) .. (1.5,0);
\draw[medarc] (3.5,0) -- (3,-0.5);
\draw[medarc] (2.5,0) -- (3,-0.5);
\draw[medarc] (3.5,0) .. controls (3,0.9) .. (2.5,0);
\draw[medarc] (4,-0.5) -- (3.5,0);
\draw[medarc] (4,-0.5) .. controls (5,0) and (4,1) .. (3.5,0);
\draw[medarc] (0,-0.5) -- (0.5,-1);
\draw[medarc] (0.5,-1) -- (0,-1.5);
\draw[medarc] (0,-1.5) .. controls (-0.9,-1) .. (0,-0.5);
\draw[medarc] (1.5,-1) -- (1,-0.5);
\draw[medarc] (1,-1.5) -- (1.5,-1);
\draw[medarc] (0.5,-1) -- (1,-1.5);
\draw[medarc] (1,-0.5) -- (0.5,-1);
\draw[medarc] (2.5,-1) -- (2,-0.5);
\draw[medarc] (2,-1.5) -- (2.5,-1);
\draw[medarc] (1.5,-1) -- (2,-1.5);
\draw[medarc] (2,-0.5) -- (1.5,-1);
\draw[medarc] (3,-0.5) -- (3.5,-1);
\draw[medarc] (3,-1.5) -- (3.5,-1);
\draw[medarc] (2.5,-1) -- (3,-1.5);
\draw[medarc] (3,-0.5) -- (2.5,-1);
\draw[medarc] (4,-1.5) .. controls (4.9,-1) .. (4,-0.5);
\draw[medarc] (3.5,-1) -- (4,-1.5);
\draw[medarc] (3.5,-1) -- (4,-0.5);
\draw[medarc] (0,-1.5) -- (0.5,-2);
\draw[medarc] (0.5,-2) -- (0,-2.5);
\draw[medarc] (0,-2.5) .. controls (-0.9,-2) .. (0,-1.5);
\draw[medarc] (1.5,-2) -- (1,-1.5);
\draw[medarc] (1,-2.5) -- (1.5,-2);
\draw[medarc] (0.5,-2) -- (1,-2.5);
\draw[medarc] (1,-1.5) -- (0.5,-2);
\draw[medarc] (2,-1.5) -- (2.5,-2);
\draw[medarc] (2.5,-2) -- (2,-2.5);
\draw[medarc] (2,-2.5) -- (1.5,-2);
\draw[medarc] (1.5,-2) -- (2,-1.5);
\draw[medarc] (3,-1.5) -- (3.5,-2);
\draw[medarc] (3.5,-2) -- (3,-2.5);
\draw[medarc] (3,-2.5) -- (2.5,-2);
\draw[medarc] (2.5,-2) -- (3,-1.5);
\draw[medarc] (4,-1.5) .. controls (4.9,-2) .. (4,-2.5);
\draw[medarc] (4,-2.5) -- (3.5,-2);
\draw[medarc] (3.5,-2) -- (4,-1.5);
\draw[medarc] (0.5,-3) -- (0,-2.5);
\draw[medarc] (0,-2.5) .. controls (-1,-3) and (0,-4) .. (0.5,-3);
\draw[medarc] (1.5,-3) -- (1,-2.5);
\draw[medarc] (0.5,-3) .. controls (1,-3.9) .. (1.5,-3);
\draw[medarc] (1,-2.5) -- (0.5,-3);
\draw[medarc] (2,-2.5) -- (2.5,-3);
\draw[medarc] (2.5,-3) .. controls (2,-3.9) .. (1.5,-3);
\draw[medarc] (1.5,-3) -- (2,-2.5);
\draw[medarc] (3.5,-3) -- (3,-2.5);
\draw[medarc] (2.5,-3) .. controls (3,-3.9) .. (3.5,-3);
\draw[medarc] (3,-2.5) -- (2.5,-3);
\draw[medarc] (3.5,-3) .. controls (4,-4) and (5,-3) .. (4,-2.5);
\draw[medarc] (4,-2.5) -- (3.5,-3);
\node[wtwo] at (0,-2.5) {};
\node[wone] at (0,-1.5) {};
\node[wone] at (0,-0.5) {};
\node[wone] at (0.5,-3) {};
\node[wtwo] at (0.5,-2) {};
\node[wtwo] at (0.5,-1) {};
\node[wone] at (0.5,0) {};
\node[wone] at (1,-2.5) {};
\node[wone] at (1,-1.5) {};
\node[wtwo] at (1,-0.5) {};
\node[wtwo] at (1.5,-3) {};
\node[wtwo] at (1.5,-2) {};
\node[wone] at (1.5,-1) {};
\node[wtwo] at (1.5,0) {};
\node[wone] at (2,-2.5) {};
\node[wtwo] at (2,-1.5) {};
\node[wone] at (2,-0.5) {};
\node[wtwo] at (2.5,-3) {};
\node[wone] at (2.5,-2) {};
\node[wone] at (2.5,-1) {};
\node[wone] at (2.5,0) {};
\node[wtwo] at (3,-2.5) {};
\node[wtwo] at (3,-1.5) {};
\node[wone] at (3,-0.5) {};
\node[wone] at (3.5,-3) {};
\node[wone] at (3.5,-2) {};
\node[wone] at (3.5,-1) {};
\node[wone] at (3.5,0) {};
\node[wtwo] at (4,-2.5) {};
\node[wtwo] at (4,-1.5) {};
\node[wone] at (4,-0.5) {};
\node[font=\small\bfseries] at (2,-4.0) {Medial orientation for the displayed face spins};
\end{tikzpicture}
\caption{On this $4\times5$ grid ($N=20$, $M=31$), the $18$ edges of $S$ (top left, teal) form four hue regions, giving $16$ shade assignments. The other $13$ edges have weight two. Each of the two upper-right squares is bounded by edges of $S$ inside one blue region, so its face spin can be chosen independently; all other faces are merged with the outer face and carry its spin. Only the outlined face has spin one in the assignment shown (top middle). The four admissible orientations each have weight $2^{13}$, and their sum is $2^{15}=2^{31-20}\cdot16$, as Lemma~\ref{four:lem:identity} predicts. The top right shows one of the $16$ shade assignments: dark and light shades alternate inside each red or blue region. Below, every corner carries one arrow, including the exterior corners. Filled amber medial vertices have weight two and open vertices have weight one.}
\label{four:fig:example}
\end{figure}
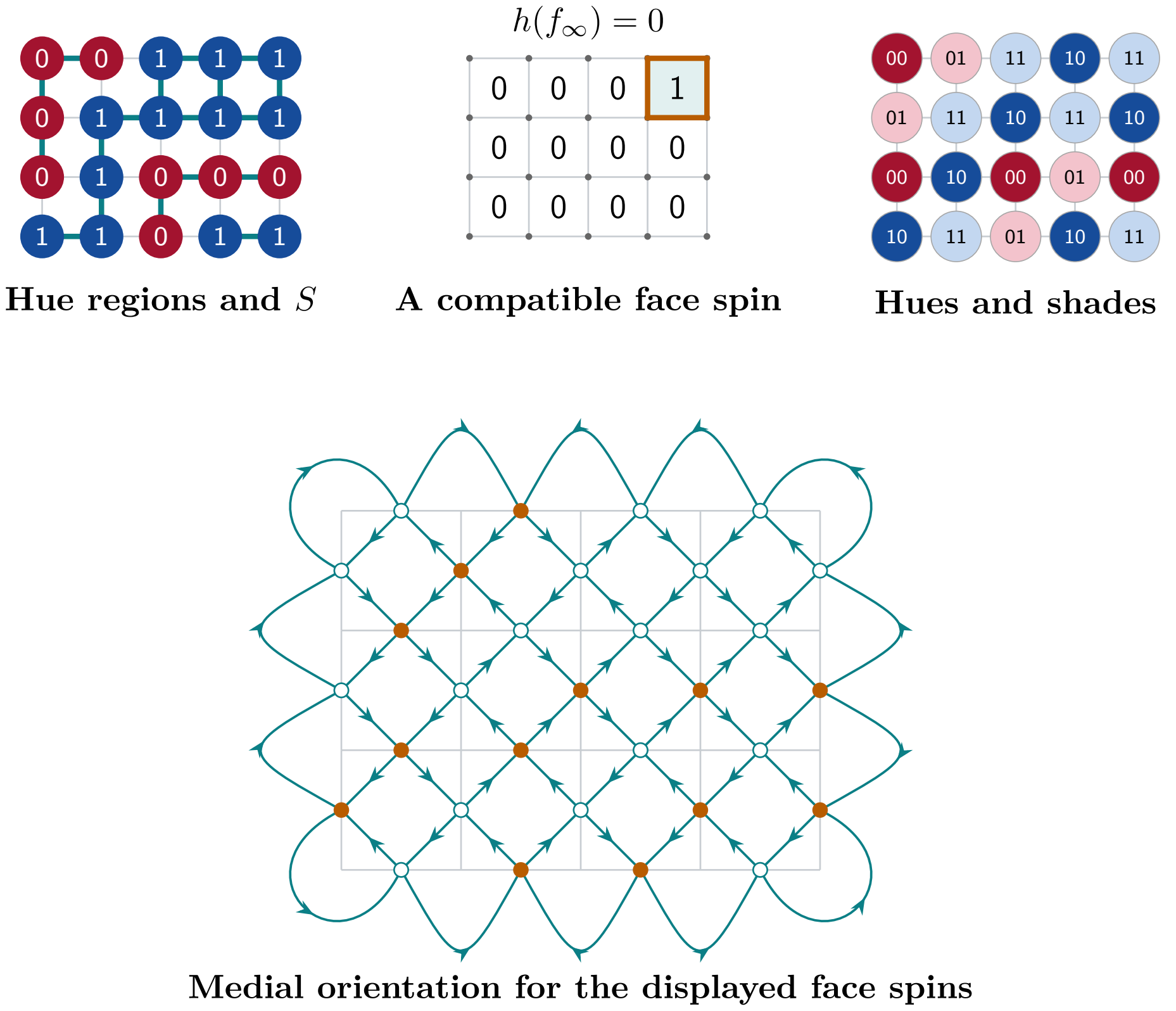

Figure~\ref{four:fig:example} illustrates Lemmas~\ref{four:lem:shades}--\ref{four:lem:identity} on a $4\times5$ grid.

% ======================================================================
\subsection{The counting algorithm}\label{four:sec:counting}

We use the following result of Chen and Liu.

\begin{theorem}[{\cite[Corollary~C.16]{CL26}}]\label{four:thm:CL}
There is an FPRAS for the six-vertex partition function $Y_6(H)$ on loopless $4$-regular multigraphs $H$, provided that at each vertex the three complementary-pair weights are nonnegative rationals, each at most the sum of the other two. The assignment of the weights to the three classes may be chosen separately at each vertex.
\end{theorem}

The weights $2,1,1$ of Figure~\ref{four:fig:types} satisfy the hypothesis, the first inequality with equality. Removing vertices of degree at most one from the primal graph enforces looplessness. This gives Algorithm~\ref{four:alg:count}.

\begin{algorithm}[t]
\caption{Approximate the number of proper four-colorings}\label{four:alg:count}
\begin{algorithmic}[1]
\Require bipartite planar graph $G$, relative error $0<\delta<1$
\State Delete duplicate edges from $G$. Set $G_0\gets G$ and $A\gets 1$.
\While{$G_0$ contains a vertex $v$ of degree at most one}
  \State Multiply $A$ by $4$ if $v$ is isolated, or by $3$ if $v$ is a leaf.
  \State Delete $v$ from $G_0$.
\EndWhile
\If{$G_0$ is empty}
  \State \Return $A$.
\EndIf
\State Retain the induced embedding, or compute a plane embedding of each component of $G_0$~\cite{BM04}.
\State Form the disjoint union $H$ of the medial graphs of the components of $G_0$.
\State At each vertex of $H$, assign the three weights of Figure~\ref{four:fig:types}.
\State Obtain $\widehat Y_6$ from the six-vertex FPRAS of Theorem~\ref{four:thm:CL} with relative error $\delta$.
\State \Return $A\,2^{|V(G_0)|-|E(G_0)|}\,\widehat Y_6$.
\end{algorithmic}
\end{algorithm}

\begin{proof}[Proof of Theorem~\ref{four:thm:main}]
We first show correctness. At each pruning step, every proper coloring of the smaller graph has exactly four extensions at an isolated vertex and exactly three at a leaf. Hence $Z_4(G)=A\,Z_4(G_0)$. If $G_0$ is nonempty, it has minimum degree at least two, so every corner joins the midpoints of two distinct edges and $H$ is a loopless $4$-regular multigraph. Put $N_0=|V(G_0)|$ and $M_0=|E(G_0)|$. Applying Lemma~\ref{four:lem:identity} to each component and multiplying gives $Z_4(G)=A\,2^{N_0-M_0}\,Y_6(H)$. This exact positive factor preserves the relative error and the success probability of the single call to the FPRAS.

\auditrevision{We now bound the running time.} Deleting duplicate edges and pruning take linear time, and so does computing a plane embedding~\cite{BM04}. To bound the cost of the six-vertex FPRAS we recall the incidence construction of Chen and Liu~\cite[proof of Corollary~C.16]{CL26}. The ground set $U$ consists of the $d=4M_0$ incidences between the $2M_0$ edges of $H$ and their endpoints. The vertex weight $\mu$ is supported on the sets that select exactly two incidences at each vertex of $H$, and multiplies their local weights. The edge weight $\nu$ is supported on the sets that select exactly one incidence of each edge of $H$, and equals one there. Both are homogeneous of degree $2M_0$, with log-concave generating polynomials~\cite[Corollary~C.16]{CL26}. A set selects one incidence per edge if and only if its complement does, so the complementary subset sum $\sum_{T\subseteq U}\mu(T)\,\nu(U\setminus T)$ is exactly $Y_6(H)$: the set $T$ records the heads of the edges in an Eulerian orientation. Sets in the support of each weight are found by arbitrary local choices, and the ratios of largest to smallest positive values are $2^{M_0}$ for $\mu$ and $1$ for $\nu$. \auditrevision{Substituting these parameters into the bound of Chen and Liu~\cite[Theorem~5.1]{CL26} gives $O(M_0^{12}\delta^{-2}\log^2(2M_0/\delta))$ arithmetic operations and value-oracle calls. An evaluation of $\mu$ or $\nu$ uses $O(M_0)$ arithmetic operations and incidence tests. Finally, a simple bipartite planar graph satisfies $M_0\le 2N_0\le 2N$ by Euler's formula~\cite{Die17}, which gives the stated arithmetic-operation bound. %For an input with parallel edges, add the work to read and deduplicate its edge list.
For  bit-complexity, the oracle values are zero, one, or powers of two with exponent at most $M_0$, and hence have polynomial encoding length and can be computed in polynomial time. Chen and Liu~\cite[Remark~A.4]{CL26} therefore give an implementation whose bit complexity is polynomial in the input length and $1/\delta$.}
\end{proof}

% ======================================================================
\subsection{Almost uniform sampling}\label{four:sec:sampling}

\begin{proof}[Proof of Corollary~\ref{four:cor:sampling}]
Run the pruning phase of Algorithm~\ref{four:alg:count} and form the weighted medial graph $H$ of the core $G_0$. If $G_0$ is empty, sample the deleted vertices as described at the end of the next paragraph.

We first show how to decode an orientation into a coloring. Given an Eulerian orientation of $H$, recover the hues $r$ on $G_0$ by the inverse map of Lemma~\ref{four:lem:spin-arrow}, fixing the outer face spin to zero in each component. This takes a traversal of the radial graph using the corner labels, as in Lemma~\ref{four:lem:spin-arrow}. Independently in each hue region, choose dark or light with probability $1/2$ at one vertex, then alternate shades throughout the region. If the orientation is drawn with probability proportional to its weight, then by the proof of Lemma~\ref{four:lem:identity} a hue assignment with $k(S)$ regions is produced with probability proportional to $2^{k(S)}$. Each of its $2^{k(S)}$ shade assignments then has conditional probability $2^{-k(S)}$, so every proper coloring of $G_0$ has the same probability. Finally, restore the deleted vertices in reverse order, choosing uniformly among the four colors at an isolated vertex or the three available colors at a leaf. Every coloring of the smaller graph has the same number of extensions, so this preserves uniformity; when $G_0$ is empty, it samples the forest $G$ directly.

It remains to sample an orientation of $H$ with probability proportional to its weight. Encode an orientation by its $2M_0$ edge directions followed by one bit per vertex of $H$, where the bit may take both values at a weight-two vertex and must be zero elsewhere. An orientation of weight $2^j$ then has exactly $2^j$ such witnesses, and a uniformly random witness projects to the desired law.

We apply the counting-to-generation theorem of Jerrum, Valiant and Vazirani~\cite[Theorem~6.3]{JVV86}. The witnesses have fixed length, can be checked in polynomial time, and are self-reducible: fixing the next bit of a padded prefix leaves an instance of the same kind. To count the completions of a prefix of edge directions, use the incidence formulation of Chen and Liu~\cite[proof of Corollary~C.16]{CL26}. The weight $\omega$ of an incidence set is the product of its incoming-pair weights if it selects two incidences at each vertex, and zero otherwise; its generating polynomial is log-concave by their proof. A partition matroid chooses one head per edge, and a prescribed direction is enforced by forbidding the opposite incidence in this matroid. The basis sum of this matroid with weight $\omega$ is exactly the number of completions of the prefix, and Lemma~C.1 of Chen and Liu~\cite{CL26} gives an FPRAS for it. Independence is tested edge by edge, any choice of two incidences per vertex gives a set with $\omega>0$, and the positive values of $\omega$ lie in $[1,2^{M_0}]$. Once all directions are fixed, the number of completions is zero for an inconsistent prefix and otherwise a power of two obtained by counting the free bits.

Finally, we bound the error. At tolerance $\eta/8$, the generator of~\cite{JVV86} succeeds with probability at least $1/2$, and given success the ratio of its largest and smallest witness probabilities is at most $(1+\eta/8)^2$. This gives total variation distance at most $\eta/2$ from the uniform distribution on witnesses. Take the first success among $\lceil\log_2(2/\eta)\rceil$ independent trials, and output the coloring $c(v)=(0,\eps(v))$ if all of them fail. Failure adds at most $\eta/2$ to the error. Discarding the witness bits and decoding cannot increase the total variation distance~\cite{LP17}, so the output is within $\eta$ of $U_G$. Each trial runs in time polynomial in the size of $G$ and $\log(1/\eta)$, and so does the whole algorithm.
\end{proof}

\section{Five-colorings of rectangles}\label{five:sec:rect-part}

This section proves Theorems~\ref{five:thm:tworow} and~\ref{five:thm:glauber} for Markov chains that uniformly sample 5-colorings.
%\subsection{Preliminaries}\label{five:sec:prelim}
We work on the $h\times w$ rectangle $P_h\,\square\,P_w$, viewed as the set $[w]\times[h]$ of sites of the square lattice $\Z^2$ with the edges joining sites at distance one, and write $n=hw$. With free boundary, no colors are prescribed outside the rectangle, and $\pi$ denotes the uniform distribution on its proper five-colorings. Our chains resample a block of sites conditional on the colors of the other sites. A neighbor outside the block either carries a prescribed color or is absent, in which case it imposes no restriction, and a sequence of prescribed colors is proper if adjacent sites in it receive different colors. Write $L(v)$ for the list of colors available at $v$ after the exclusions by its prescribed neighbors. As in Section~\ref{sec:intro}, $0<\varepsilon<1/2$ and mixing times are measured from the worst initial coloring. We write $D_\varepsilon=n+\log(1/\varepsilon)$.

A \emph{heat bath} on a block resamples its sites from the uniform distribution conditional on the colors outside the block. Every chain discussed below has exactly uniform stationary law, and its states are proper colorings. Greedy coloring supplies a starting coloring, since the maximum degree is four.

For a reversible kernel $P$ write
\[
  \cE_P(f)=\tfrac12\sum_{x,y}\pi(x)P(x,y)\bigl(f(x)-f(y)\bigr)^2 .
\]
An inequality $\Var_\pi(f)\le a\,\cE_P(f)$ gives inverse spectral gap at most $a$. A heat bath is a conditional averaging projection for each chosen block. A one-site color proposal on a fiber with $b$ legal colors has the form $(1-b/5)I+(b/5)U$, with $U$ conditional averaging. These kernels have nonnegative spectrum. Since $\pi_{\min}\ge5^{-n}$, the reversible spectral estimate gives $\tmix(\varepsilon)=O(aD_\varepsilon)$ for them~\cite[Chapter~12]{LP17}. We also use the coupling inequality and the triangle inequality for the minimum expected Hamming distance~\cite[Chapter~4]{LP17}. Interpolating a sequence of boundary prescriptions uses the gluing principle familiar from path coupling~\cite{BD97}, with all intermediate conditional laws specified where they are used.

Two facts let every local chain run on boundedly many fair bits. A uniform choice among $m$ labeled moves can be padded to the next power of two by adding holding slots; conditional on selecting a real slot the choice is uniform, and real slots have probability greater than $1/2$ (or one). For a block with $Z$ legal completions, draw a uniform integer $J$ with $\lceil\log_2Z\rceil+b$ bits. When $J<Z\lfloor 2^{\lceil\log_2Z\rceil+b}/Z\rfloor$, install the completion of rank $J\bmod Z$; otherwise hold. The filling error from the ideal heat bath is at most $2^{-b}$ in total variation, and the kernel remains uniform and reversible on the fiber.

% ======================================================================
\subsection{A uniform estimate for two-row strips}\label{five:sec:strip}

Consider the conditional law on a $2\times w$ strip. Each of its two exterior rows is either absent or carries a full proper sequence of prescribed colors. For the comparison below, a changing exterior row can temporarily have an equality across the cut between its old and new portions. Only inequalities between a strip site and its exterior neighbors enter its conditional distribution. These intermediate distributions are well defined, because each strip site has at most four neighbors including its pins, so a greedy assignment always extends the given restrictions. Write $W_d$ for the minimum expected cost $d$ among couplings of two laws. We estimate the transmission of one exterior disagreement through a block, as in the block and recursive couplings of~\cite{AMMV04,GMP05}, for strips of unbounded length.

\begin{lemma}[Two-row strip estimate]\label{five:lem:strip}
Consider a $2\times w$ strip whose two ends are free. Change the top exterior color at column $k$ from $s$ to $t\ne s$. On each side of $k$ the top exterior sequence is proper, and, when they exist, its immediate left neighbor differs from $t$ and its immediate right neighbor differs from $s$. The bottom exterior row is proper or absent. Then the laws $\mu_s,\mu_t$ of the strip before and after the change satisfy
\[
  W_\Ham(\mu_s,\mu_t)\le c,\qquad c=.999970.
\]
This includes changes at the end columns and $w=1$, and exchanging the two strip rows gives the same estimate for a change in the bottom exterior row.
\end{lemma}

\begin{proof}
We first set up a recurrence along the strip. The $20$ column states form $\cS=\{(a,b):0\le a,b<5,\ a\ne b\}$. Let $H_{ij}$ be their Hamming distance and set $T_{ij}=\one_{\{H_{ij}=2\}}$. For an upper and lower exterior pair $p=(u,v)$, the diagonal matrix $D_p$ has entries $\one_{\{i_1\ne u,\,i_2\ne v\}}$; we write $v=5$ when the bottom wall is absent. A word $w'=p_1\cdots p_m$ records the exterior pairs of a half-strip in order away from the changed column. For a fixed physical end, let $F_{w'}(i,j)$ be the minimum expected suffix Hamming distance between the conditional laws for incoming states $i,j$. A nonempty suffix admits every incoming state: even at the last column, the preceding column, the wall and the physical end together exclude at most three colors in each coordinate, leaving an unequal pair available. The terminal cost is zero, and incoming states incompatible with a physical end receive zero mass when that end is reached. Let $v$ denote the vector of numbers of continuations after the first column. Coupling the first columns and then their remaining conditional suffixes gives
\begin{equation}\label{five:eq:recurrence}
  k_i(z)=\frac{T_{iz}D_p(z,z)v_z}{(TD_pv)_i},\qquad F_{pw'}(i,j)\le W_{H+F_{w'}}(k_i,k_j).
\end{equation}
The distances $F_{w'}$ are pseudometrics on incoming states.

{\color{changes}The recurrence reduces Lemma~\ref{five:lem:strip} to a finite computation once two difficulties are handled: strips have unbounded length, and the laws $k_i$ depend on counts of continuations that grow with the length. Both are handled by enclosure. When a half-strip is specified for only a few columns, every possible vector of continuation counts lies in a cone spanned by finitely many vectors, so the laws $k_i$ lie in boxes computed from these vectors. A table of upper bounds on $F_{w'}$, indexed by the first two exterior pairs of the half-strip, is then checked to be consistent with~\eqref{five:eq:recurrence}, and induction on the length bounds $F_{w'}$ for half-strips of every length. Finally the two laws of the changed column are coupled using these bounds on both sides, and a finite case analysis over the exterior pairs nearest the changed column shows that every case costs less than $c$. All quantities are computed in integer arithmetic and rounded outward, so every accepted number is a rigorous upper bound. Appendix~\ref{app:certificate} completes the proof. The computation is carried out by a short C++ program, available as ancillary files with the arXiv version of this paper, which runs in about nine minutes on a laptop.}
\end{proof}

% ======================================================================
\subsection{The two-row chain on rectangles}\label{five:sec:rect}

\begin{proof}[Proof of Theorem~\ref{five:thm:tworow}]
Number the rows $1,\dots,h$ and let the strip $B_j$ consist of rows $j$ and $j+1$; its exterior rows are rows $j-1$ and $j+2$ when they exist, and its ends are free. For any two initial colorings and a common strip choice, change the top exterior row of the strip from one coloring to the other from left to right, and then the bottom row from left to right. At each change the currently installed left portion agrees with the new proper sequence and the right portion with the old one, and the opposite row is proper or absent throughout. Lemma~\ref{five:lem:strip} and the triangle inequality for transport therefore bound the expected number of discrepancies in the resampled strip by $c$ times the number of changed exterior sites. Each row $i$ belongs to $b_i\ge1$ strips and borders at most $b_i$ strips. Writing $m=h-1$, coupling common strip choices therefore contracts Hamming distance for every pair of colorings by a factor at most $1-(1-c)/m$. The Lipschitz seminorm of any nonconstant eigenfunction contracts by the same factor, which gives the stated gap. Each heat bath preserves the uniform law on all colorings.

For bounded random bits, choose a strip by padding the $m$ choices to $2^{\lceil\log_2m\rceil}$ slots and holding on unused slots. On a real choice, count the $Z\le20^w$ fillings by the column transfer recurrence and use suffix counts to unrank. Draw a uniform integer $J$ with $b_*=\lceil\log_2Z\rceil+\lceil\log_2(10^6nm)\rceil$ bits. Install the filling of rank $J\bmod Z$ when $J<Z\lfloor2^{b_*}/Z\rfloor$, and hold otherwise. Each legal filling has the same probability, and the holding probability depends only on the exterior, which gives reversibility. Relative to an ideal draw, coupling costs at most $n/(10^6nm)$ in expected Hamming distance per starting coloring. For a pair of distinct colorings this adds at most $2\cdot10^{-6}/m$ times their initial Hamming distance. The real-slot probability is at least $1/2$, so the net contraction per step is at most $1-1/(10^5m)$. Coupling to stationarity from distance at most $n$ gives the claimed mixing time, and hence irreducibility. Only $20$ column states and $O(w)$-bit counts enter the update. For $h=1$, count list-path colorings by the one-dimensional recurrence; choosing $\lceil\log_2Z\rceil$ bits and installing a uniformly ranked path on a successful draw refreshes with probability at least $1/2$, which gives the asserted bound.
\end{proof}

% ======================================================================
\subsection{Glauber dynamics on rectangles}\label{five:sec:local}

We now localize the two-row chain to single sites. Orient the rectangle so that its rows have length $w$, and let $G$ be Glauber dynamics: choose a uniform free vertex, propose a uniform one of the five colors, and accept a legal proposal. We prove an inverse spectral gap of $O(nw^3)$ and then take $w=s$.

\begin{proof}[Proof of Theorem~\ref{five:thm:glauber}]
We use the path comparison of Diaconis and Saloff-Coste~\cite{DSC93}. For an oriented transition $e$ of the simulating chain with stationary edge weight $q(e)$, bound the congestion by
\[
  q(e)^{-1}\sum_{\gamma\ni e}\pi(x)P(x,y)\,|\gamma| .
\]
Here $\gamma$ is the chosen route for a transition $x\to y$ of the simulated kernel $P$, and each traversal is counted separately. By Cauchy--Schwarz the maximum congestion bounds the ratio of the two Dirichlet forms. Fix deterministic tie-breaking and write $Q$ for the number of completions of a fixed two-row strip exterior, so that all conditional stationary masses equal $1/Q$.

We first compare a full strip refresh with updates of single sites and horizontal pairs. Let $D$ choose uniformly among all strip singletons and horizontal neighboring pairs, and resample the chosen block conditionally; there are $4w-2$ choices. Route a full refresh $X\to Y$ by fixing vertices to their target colors in column order, top before bottom. At a pivot $v$ with target $a=Y(v)$, clear each unprocessed neighbor $u$ currently colored $a$. If $u$ has a right neighbor, update the horizontal pair consisting of $u$ and that neighbor. Each site in the pair has at least two colors legal against all its other neighbors, because at most three neighbors altogether remain after the pair edge is excluded; choose a color other than $a$ at $u$, and then a distinct legal color at its mate. At the right endpoint use a singleton: since the right side is free, its total degree is at most three and at least two colors are available. No such mate is an additional neighbor of $v$, and no processed site is touched. Now set $v=a$. This constructs a route of length $O(w)$.

While column $j$ is being processed, the current strip agrees with $Y$ to the left of $j$ and with $X$ beyond $j+2$. For any fixed directed transition, choose one site it actually changes. Its column lies in $[j,j+2]$, so there are at most three possible pivot columns $j$. Store the constant-sized substep, both endpoint assignments on columns $j$ through $\min(j+2,w)$, and one complementary proper strip coloring. To construct the latter, copy $X$ to the left of $j$ and $Y$ to its right and fill column $j$; each of its two sites has at least two legal colors before their mutual edge is tested, so one can choose unequal colors there. The recorded strip coloring and the state before the directed transition recover both $X$ and $Y$ off the recorded columns. This encoding bounds the number of routed occurrences through that transition by $O(Q)$. The refresh weight is $Q^{-2}$, and a used $D$-edge has weight at least $1/(25(4w-2)Q)$. Thus
\begin{equation}\label{five:eq:refresh-pairs}
  \Var(f)\le O(w^2)\,\cE_D(f).
\end{equation}

Next we simulate a pair update by single-site moves. When the other row and the exterior are held fixed, a row is a list-colored path. Before its horizontal neighbors are tested, each site has at least three colors, since it has at most two other neighbors. Suppose a labeled $D$-move changes sites $k,k+1$. After testing their horizontal neighbors outside the pair, let their lists be $A,B$; both have size at least two. The possible assignments are the edges of the bipartite graph with parts $A,B$ and edges $(u,v)$ for $u\ne v$, and incident edges differ at one site. If one part has size at least three, any two vertices in the other share a neighbor, and all vertices have positive degree. When both parts have size two with different color sets, at most one edge of $K_{2,2}$ is missing. Thus, except when $A=B=\{a,b\}$, any pair update has a deterministic single-site route of at most $24$ steps.

In the exceptional case write the required change as $(a,b)\to(b,a)$. Position $k+2$ exists, because position $k+1$ has at least three list colors before its right neighbor is tested and the right side is free. The current color $c$ at $k+2$ is distinct from $a,b$, and the list at $k+1$ before its right neighbor is tested is exactly $\{a,b,c\}$. At $k+2$ choose a list color $t$ avoiding $c$ and the color at $k+3$, if present, and execute
\begin{equation}\label{five:eq:swap}
\begin{aligned}
  (b,c)&\to(c,t)&&\text{on }(k+1,k+2),\\
  a&\to b&&\text{at }k,\\
  (c,t)&\to(a,c)&&\text{on }(k+1,k+2).
\end{aligned}
\end{equation}
All three endpoint changes are legal. An exceptional child move has to swap its two input colors, so the first child can be exceptional only for $t=b$ and the last only for $t=a$. At most one of them recurses, one site farther to the right, so the route has length $O(w)$.

One can bound the congestion without recording the recursion stack. Fix an oriented single-site move with pre-state $Z$; the row and the changed column $i$ are known. Record the root pair index $j$ and its target colors. For an ordinary root, record the two root input colors and the substep. For an exceptional root, let $\ell$ be the deepest exceptional call active at this move. The move is its own pivot or a move of an ordinary child, so $\ell\in\{i-2,i-1,i\}$. Record $\ell$, its three input colors, and the substep in its nonrecursive work (at most $49$ choices). Restoring those three colors in $Z$ recovers the state at entry to this frame, because any child that already returned changed only its two specified sites. The active exceptional ancestors occupy the consecutive starts $\ell-1,\dots,j$. At each such index $k$, the visible color at $k$ is $g$ and the input pair of its child is $(s_1,t_1)$. The parent list $A=\{a,b\}$ is computable from the unchanged other row and the visible left neighbor. If its exceptional child was the first move in~\eqref{five:eq:swap}, then $(g,s_1,t_1)=(a,b,c)$ with $s_1\in A$ and $t_1\notin A$; if it was the last, then $(g,s_1,t_1)=(b,c,a)$ with $t_1\in A$ and $s_1\notin A$. In the latter case restore the parent input as $(t_1,g,s_1)$; in the former the visible triple is already its input. This reconstructs the original labeled pair move, and the deterministic routes and the recorded substep distinguish its occurrences. Only $j$ ranges over $w$ choices, so $O(w)$ routed occurrences use the fixed single-site move. Singletons route directly.

Let $S$ denote the random-site, random-color chain in the strip. A used $S$-edge has stationary weight $1/(10wQ)$, and a labeled $D$-move has weight at most $1/((4w-2)Q)$. The $O(w)$ route length and the $O(w)$ occurrence bound give $\cE_D(f)\le O(w^2)\cE_S(f)$, and by~\eqref{five:eq:refresh-pairs} the conditional inverse gap of a strip is $O(w^4)$.

It remains to average over strips. For $h\ge2$, the ideal two-row chain has inverse gap $O(h)$ by Theorem~\ref{five:thm:tworow}, and its Dirichlet form averages the strip variances. After the conditional bound is applied, each of the $n$ sites occurs in at most two of the $h-1$ strips, each giving it update probability $1/(2w)$. This yields
\[
  \Var_\pi(f)\le O(h)\,O(w^4)\,\frac{h}{h-1}\,\cE_G(f)=O(nw^3)\,\cE_G(f).
\]
At width one the two-site strip uses only singletons, and the first comparison applies with $w=1$. When $h=1$, use the same refresh-to-horizontal-pair sweep on the single row. Only the right neighbor can block; the pair endpoints retain two options, and a right-end singleton has at least two. At bridge position $j$ one can fill greedily, since its total degree, including pins, is at most four. The same occurrence encoding gives $O(w^2)$ for this comparison, and the list-path argument above gives the second factor $O(w^2)$, so the inverse gap is $O(w^4)=O(nw^3)$ in this case too.

All moves are reversible for the uniform distribution. A one-site proposal on a fiber with $b$ permitted colors acts as $(1-b/5)I+(b/5)U$, with $U$ the uniform fiber averaging operator, so the kernel is positive semidefinite. A positive Poincar\'e bound gives irreducibility, and the spectral mixing estimate with $\pi_{\min}\ge5^{-n}$ gives the time bound. With $w=s$ this is $O(ns^3)$.

On a path one can instead use a single-site coupling directly, as in the high-color regime~\cite{Jer95}. For a site with neighbor-color set $F$ and current color $z$, the proposal law is $\frac15\sum_{a\notin F}\delta_a+|F|\delta_z/5$. Changing $z$ alone costs at most $2/5$ in total variation, and changing the color of one neighbor costs at most $1/5$; these estimates still hold for intermediate assignments, by the displayed law. In a coupled random update, a differing site therefore loses at least $3/5$ of its discrepancy when it is chosen, and supplies at most $1/5$ when each of its at most two neighbors is chosen. Expected Hamming distance contracts by $1-1/(5n)$, and coupling to stationarity gives $O(n\log(n/\varepsilon))$ steps. If a worst-case bound on fair bits is required, choose one of $2^{\lceil\log_2(5n)\rceil}$ slots and hold on those not used by the $5n$ proposals; this multiplies all off-diagonal rates by the same factor of at least $1/2$.
\end{proof}

% ======================================================================
\section{Conclusions}\label{sec:remarks}

The four-color algorithm uses a bipartition and a plane embedding, but no rectangular shape, no degree bound and no face-length condition. A hole in a grid domain is just another face of the plane graph, and any face may serve as $\finf$. Bipartiteness enters only in Lemma~\ref{four:lem:shades}, where the shades alternate along same-hue edges, and planarity enters through the face spins and the medial graph.

The five-color proofs use the two-row strips only through Lemma~\ref{five:lem:strip}. Everything else, namely the interpolation, the bookkeeping, the implementation and the comparisons with local dynamics, is elementary. The strip estimate is the one place where five colors and the geometry of the square lattice enter in an essential way, and the one place where a computer is used.

The two methods are complementary. The four-color reduction is exact and global: it relies on planarity and bipartiteness and analyzes no local chain. The five-color argument is local and computer-assisted, and it relies on the geometry of the square lattice. Neither covers the other case, since the hue--shade decomposition is specific to four colors on bipartite graphs and the strip certificate is a finite computation for five colors.

Several questions remain. We do not know whether Glauber dynamics for four-colorings of rectangular grids with free boundary is rapidly mixing; our algorithm sidesteps this question but does not answer it. \auditrevision{The exponent $13$ in the arithmetic-operation bound of Theorem~\ref{four:thm:main} comes from the general bound of Chen and Liu together with the cost of evaluating the incidence weights; a direct analysis of this special case may give a much smaller exponent.} Four-colorings of non-bipartite planar graphs, such as domains in the triangular lattice, are not covered by the reduction. For five colors, our bound for Glauber dynamics on rectangles is polynomial, but for six or more colors strong spatial mixing gives $O(n\log n)$ mixing for block dynamics~\cite{DSVW04,AMMV04}, and it would be interesting to know whether the same holds for five colors. We also do not know whether the uniform five-coloring of the square lattice has strong spatial mixing.

% ======================================================================

\appendix
\section{The certificate for Lemma~\ref{five:lem:strip}}\label{app:certificate}

{\color{changes}This appendix completes the proof of Lemma~\ref{five:lem:strip}, continuing from the recurrence~\eqref{five:eq:recurrence} and the notation of Section~\ref{five:sec:strip}.}

\subsection{Unspecified continuations}\label{app:anchors}

The recurrence must handle strips of every length, so we represent continuations that are not specified. After a specified prefix $p_1,\dots,p_m$, use the matrix
\[
  A=(TD_{p_1})\cdots(TD_{p_m})E .
\]
{\color{changes}At a physical end, which is free, all columns of $E$ equal $\one$.} For an unspecified continuation, take $E=T$. Every possible suffix-count message belongs to the nonnegative cone of the columns of $A$, by conditioning on the next column. {\color{changes}This enclosure also includes a free end immediately after the prefix, since $T\one=13\one$.} We call the generating columns of this cone \emph{anchors}.

\subsection{Bounds for half-strips}\label{app:half}

For use in~\eqref{five:eq:recurrence} the checker builds upper bounds $U_{p,q}(i,j)$ indexed by the first two pin pairs. With two exterior walls put $K=5$, and with one put $K=4$. For every proper word of length $2,\dots,K-1$, it applies~\eqref{five:eq:recurrence} back from {\color{changes}a free physical end}. For every proper prefix of length $K$, it encloses the possible first-column laws using $A$ for the last $K-1$ letters and uses the cost $H+U_{p_2,p_3}$. Within each $(p_1,p_2)$ class it takes entrywise maxima over cases and column-state relabelings, followed by the shortest-path closure of the distance matrix. Every true $F_{w'}$ obeys the triangle inequality, so replacing a bound by the minimum of path sums preserves its validity. Starting from zero, the program generates a candidate $U$ and verifies that this complete update is entrywise at most $U$. {\color{changes}It obtains the candidate by iterating the update from the zero table, enlarging all entries slightly every sixth round, and stops as soon as the update of the current table is entrywise at most that table.} Induction on the true word length then proves $F_{w'}\le U_{p_1,p_2}$ for every word with at least two letters. Shorter words are evaluated directly. {\color{changes}The two-wall and one-wall checks use $13{,}846$ and $5$ length-$K$ prefixes up to color permutation, respectively, and the largest entries of the certified tables are $1.587253$ and $1.358039$.}

\subsection{Transport between enclosed laws}\label{app:transport}

It remains to bound transport when the continuation message is known only to lie in this cone. For an arbitrary continuation $v=A\gamma$ with $\gamma\ge0$, define $B_i(z)=\sum_\ell T_{i\ell}D_p(\ell,\ell)A_{\ell z}$. Each $k_i(\ell)$ lies between the extrema over $z$ of $T_{i\ell}D_p(\ell,\ell)A_{\ell z}/B_i(z)$. Indeed, the ratio for a mixture is a convex combination of the anchor ratios, with weights proportional to $B_i(z)\gamma_z$. On the common support,
\[
  k_i(\ell)-k_j(\ell)=k_i(\ell)\left(1-\frac{\sum_zB_i(z)\gamma_z}{\sum_zB_j(z)\gamma_z}\right),
\]
and the same ratio rule bounds the last fraction. Off the common support only one signed term contributes. These formulas enclose every coordinate of $k_i-k_j$. For a metric cost matrix $M$ of diameter $D$, probabilities $p,q$ and nominal probabilities $p_0,q_0$ satisfy
\begin{equation}\label{five:eq:transport}
  W_M(p,q)\le W_M(p_0,q_0)+\tfrac12D\,\bigl\|(p-q)-(p_0-q_0)\bigr\|_1 .
\end{equation}
The signed residual has positive and negative parts of equal mass; sending this mass across distance at most $D$ and composing flows gives the inequality. The checker rounds probability intervals outward in units of $10^{-7}$, constructs a feasible integral transport for the nominal distributions, and bounds the residual by summing the coordinate errors. It rounds costs up in units of $10^{-6}$. Thus every accepted cost is an upper bound even if its nominal transport is suboptimal.

\subsection{The changed column}\label{app:center}

Finally we couple the changed column. Relabel $s=0$, $t=1$, and denote the opposite central pin by $b$, taking $b=5$ for an absent row. For left and right matrices $A_L,A_R$ as above, set
\[
  w_i=(A_L\gamma)_i(A_R\eta)_i\one_{\{i_2\ne b\}},\qquad p_i=\frac{w_i\one_{\{i_1\ne0\}}}{Z_0},\qquad q_i=\frac{w_i\one_{\{i_1\ne1\}}}{Z_1},
\]
where $Z_0,Z_1$ normalize the distributions. Given the central column, the two sides are independent. For suffix upper bounds $S_L,S_R$, a feasible upper cost for coupling two center states is
\[
  M(i,j)=H_{ij}+S_L(i,j)+S_R(i,j).
\]
The first check normalizes each half message to sum to one, takes coordinate extrema of its normalized anchors, and multiplies the two intervals for each $w_i$. On the common support $p_i-q_i=p_i(Z_1-Z_0)/Z_1$, with $Z_1-Z_0=\sum_{i_1=0}w_i-\sum_{i_1=1}w_i$. Signed coordinate intervals and~\eqref{five:eq:transport} give a direct transport bound. A second bound conditions on a pair $a$ of far anchors. Let $p_a,q_a$ be the resulting normalized center laws. The mixing weights on anchors under the first law are those under the second, tilted by $g_a=Z_{0,a}/Z_{1,a}$. For $\ell\le g_a\le u$ with $\ell>0$, the total variation distance between these weights is at most $\min\{1,(u-\ell)/(4\ell)\}$, since the variation is $\E|g-\E g|/(2\E g)$ and $\Var g\le(u-\ell)^2/4$. Applying a fixed-anchor transport and coupling the residual mixture of the $q_a$'s gives the upper bound
\[
  \max_aW_M(p_a,q_a)+\diam(M)\min\{1,(u-\ell)/(4\ell)\}\min\Bigl\{1,\tfrac12\sum_i\bigl(\max_aq_a(i)-\min_aq_a(i)\bigr)\Bigr\}.
\]
The last factor bounds the total variation distance between any two $q_a$. Both far ends are treated independently.

The central enumeration that makes up the finite certificate specifies zero, one or two pin pairs on each side. {\color{changes}A side with fewer than two pairs ends physically, at a free end.} A side with two pairs initially represents every continuation. Each unresolved node splits on one side into {\color{changes}its free physical end} and every proper next pair, and the bounds from~\eqref{five:eq:recurrence} are tightened as more pairs become known. {\color{changes}Nodes are refined until each side has ended or carries five pairs; a node whose bound is still too large at that point would make the check fail, and none does.} Thus every finite strip is represented at every refinement. Relabeling colors, or reversing the columns and exchanging old and new, leaves the three cases in the table below. {\color{changes}The counts are canonical root classes, before refinement.} Only nodes with upper bound strictly below $999970$ in distance units of $10^{-6}$ are accepted.
\begin{center}\small
\begin{tabular}{@{}lll@{}}
\toprule
Opposite central pin & Root classes & Largest accepted upper bound\\
\midrule
$0$ or $1$ & $12{,}645$ & {\color{changes}$0.999965$}\\
$2$, $3$ or $4$ & $37{,}485$ & {\color{changes}$0.999962$}\\
Absent & $101$ & {\color{changes}$0.934373$}\\
\bottomrule
\end{tabular}
\end{center}
The inductive inequalities make this finite computation apply to all widths. {\color{changes}This completes the proof of Lemma~\ref{five:lem:strip}.}

\subsection{The verifier}\label{app:verifier}

{\color{changes}The \auditrevision{core }verifier consists of three C++ files and a shell script, distributed as ancillary files with the arXiv version of this paper. The header \texttt{transfer.hpp} contains the exact arithmetic: the column states and transfer matrices, outward rounding of probabilities in units of $10^{-7}$ and of costs in units of $10^{-6}$, a successive-shortest-path solver that returns an integral transport between two rounded distributions, and the enclosures of the laws $k_i$ from Section~\ref{app:transport}. The program \texttt{half\_strip.cpp} builds and certifies the tables $U$ of Section~\ref{app:half}; it is run once with two walls ($K=5$) and once with one ($K=4$), and writes each table to a file. The program \texttt{center.cpp} reads a table and runs the enumeration of Section~\ref{app:center} for one class of opposite central pins, using the direct transport bound first and the anchor bound when the direct one does not suffice. The script \texttt{check.sh} compiles both programs with assertions enabled and signed overflow trapped, builds both tables from zero, and runs the three central checks, each of which ends with a line beginning \texttt{PASS}. A failed assertion, an overflow or a case that cannot be certified stops the script with a nonzero exit status. With Apple clang~21 on an Apple M3 laptop the complete run takes about nine minutes.}

\auditrevision{The complete verification consists of certifying both half-strip tables and then checking all three central classes. The central program assumes that its input table has already been certified; its \texttt{PASS} message alone does not validate an arbitrary table. The script \texttt{check.sh} performs the complete workflow and uses the unsharded central checks.

The optional arguments to the central program specify an offset \texttt{off} and a stride \texttt{stride} for splitting the canonical root cases. They must satisfy $\texttt{stride}>0$ and $0\le\texttt{off}<\texttt{stride}$. The standalone executable assumes valid shard arguments. A complete sharded check uses every offset $0,\dots,\texttt{stride}-1$ with the same stride and certified table, requires every run to succeed, and checks that the combined root count equals the corresponding count in Section~\ref{app:center}. The ancillary driver \texttt{check\_sharded.py} validates the shard arguments and enforces these coverage checks for the class $b=2$, including the expected root count for each individual shard.

We completed the sharded verification of the class $b=2$, representing opposite central pins $2$, $3$, or $4$, with \texttt{check\_sharded.py}, which certified the two-wall table from zero and ran all eight offsets with stride $8$. Every shard exited successfully: offsets $0,\dots,4$ each checked $4{,}686$ canonical roots, and offsets $5,6,7$ each checked $4{,}685$, for a total of $37{,}485$. The largest accepted upper bound was $999962/10^6=0.999962<0.999970$, agreeing with the unsharded check and the table in Section~\ref{app:center}. Thus the reported run includes argument validation and verification of the entire central class. The file \texttt{README.txt} records the results.}

\end{document}

\subsection{The verifier}\label{app:verifier}

{\color{changes}The \auditrevision{core }verifier consists of three C++ files and a shell script, distributed as ancillary files with the arXiv version of this paper. The header \texttt{transfer.hpp} contains the exact arithmetic: the column states and transfer matrices, outward rounding of probabilities in units of $10^{-7}$ and of costs in units of $10^{-6}$, a successive-shortest-path solver that returns an integral transport between two rounded distributions, and the enclosures of the laws $k_i$ from Section~\ref{app:transport}. The program \texttt{half\_strip.cpp} builds and certifies the tables $U$ of Section~\ref{app:half}; it is run once with two walls ($K=5$) and once with one ($K=4$), and writes each table to a file. The program \texttt{center.cpp} reads a table and runs the enumeration of Section~\ref{app:center} for one class of opposite central pins, using the direct transport bound first and the anchor bound when the direct one does not suffice. The script \texttt{check.sh} compiles both programs with assertions enabled and signed overflow trapped, builds both tables from zero, and runs the three central checks, each of which ends with a line beginning \texttt{PASS}. A failed assertion, an overflow or a case that cannot be certified stops the script with a nonzero exit status. With Apple clang~21 on an Apple M3 laptop the complete run takes about nine minutes.}

\auditrevision{The complete verification consists of certifying both half-strip tables and then checking all three central classes. The central program assumes that its input table has already been certified; its \texttt{PASS} message alone does not validate an arbitrary table. The script \texttt{check.sh} performs the complete workflow and uses the unsharded central checks.

The optional arguments to the central program specify an offset \texttt{off} and a stride \texttt{stride} for splitting the canonical root cases. They must satisfy $\texttt{stride}>0$ and $0\le\texttt{off}<\texttt{stride}$. The current standalone program does not validate these arguments, so they must be checked before invocation. For a complete sharded check of a central class, run every offset $0,\dots,\texttt{stride}-1$ with the same stride and certified table, require every run to succeed, and verify that the combined root count equals the corresponding count in Section~\ref{app:center}. A success for one shard certifies only that shard.

We have completed this sharded verification for the class $b=2$, representing opposite central pins $2$, $3$, or $4$, using all eight offsets with stride $8$ and the certified two-wall table. Every shard exited successfully: offsets $0,\dots,4$ each checked $4{,}686$ canonical roots, and offsets $5,6,7$ each checked $4{,}685$, for a total of $37{,}485$. The largest accepted upper bound was $999962/10^6=0.999962<0.999970$, agreeing with the unsharded check and the table in Section~\ref{app:center}. The ancillary script \texttt{check\_sharded.py} reproduces this verification by certifying the two-wall table and then running all shards; \texttt{README\_verification.txt} records the results.}

\end{document}

\subsection{The verifier}\label{app:verifier}

{\color{changes}The verifier consists of three C++ files and a shell script, distributed as ancillary files with the arXiv version of this paper. The header \texttt{transfer.hpp} contains the exact arithmetic: the column states and transfer matrices, outward rounding of probabilities in units of $10^{-7}$ and of costs in units of $10^{-6}$, a successive-shortest-path solver that returns an integral transport between two rounded distributions, and the enclosures of the laws $k_i$ from Section~\ref{app:transport}. The program \texttt{half\_strip.cpp} builds and certifies the tables $U$ of Section~\ref{app:half}; it is run once with two walls ($K=5$) and once with one ($K=4$), and writes each table to a file. The program \texttt{center.cpp} reads a table and runs the enumeration of Section~\ref{app:center} for one class of opposite central pins, using the direct transport bound first and the anchor bound when the direct one does not suffice. The script \texttt{check.sh} compiles both programs with assertions enabled and signed overflow trapped, builds both tables from zero, and runs the three central checks, each of which ends with a line beginning \texttt{PASS}. A failed assertion, an overflow or a case that cannot be certified stops the script with a nonzero exit status. With Apple clang~21 on an Apple M3 laptop the complete run takes about nine minutes.}

\end{document}

\subsection{The verifier}\label{app:verifier}

{\color{changes}The verifier consists of three C++ files and a shell script, distributed as ancillary files with the arXiv version of this paper. The header \texttt{transfer.hpp} contains the exact arithmetic: the column states and transfer matrices, outward rounding of probabilities in units of $10^{-7}$ and of costs in units of $10^{-6}$, a successive-shortest-path solver that returns an integral transport between two rounded distributions, and the enclosures of the laws $k_i$ from Section~\ref{app:transport}. The program \texttt{half\_strip.cpp} builds and certifies the tables $U$ of Section~\ref{app:half}; it is run once with two walls ($K=5$) and once with one ($K=4$), and writes each table to a file. The program \texttt{center.cpp} reads a table and runs the enumeration of Section~\ref{app:center} for one class of opposite central pins, using the direct transport bound first and the anchor bound when the direct one does not suffice. The script \texttt{check.sh} compiles both programs with assertions enabled and signed overflow trapped, builds both tables from zero, and runs the three central checks, each of which ends with a line beginning \texttt{PASS}. A failed assertion, an overflow or a case that cannot be certified stops the script with a nonzero exit status. With Apple clang~21 on an Apple M3 laptop the complete run takes about nine minutes.}

\auditrevision{The complete verification consists of certifying both half-strip tables and then checking all three central classes. The central program assumes that its input table has already been certified; its \texttt{PASS} message alone does not validate an arbitrary table. The script \texttt{check.sh} performs the complete workflow and uses the unsharded central checks.

The optional arguments to the central program specify an offset \texttt{off} and a stride \texttt{stride} for splitting the canonical root cases. They must satisfy $\texttt{stride}>0$ and $0\le\texttt{off}<\texttt{stride}$. The current standalone program does not validate these arguments, so they must be checked before invocation. For a complete sharded check of a central class, run every offset $0,\dots,\texttt{stride}-1$ with the same stride and certified table, require every run to succeed, and verify that the combined root count equals the corresponding count in Section~\ref{app:center}. A success for one shard certifies only that shard.}

\end{document}